\documentclass[12pt]{article}
\usepackage{a4wide,epsfig,feynarts}
\newcommand{\agt}{\,\rlap{\lower 3.5 pt \hbox{$\mathchar \sim$}} \raise 1pt
 \hbox {$>$}\,}
\newcommand{\alt}{\,\rlap{\lower 3.5 pt \hbox{$\mathchar \sim$}} \raise 1pt
 \hbox {$<$}\,}

\begin{document}

\title{
\vskip-3cm{\baselineskip14pt
\centerline{\normalsize DESY 04-108\hfill ISSN 0418-9833}
\centerline{\normalsize hep-ph/0406254\hfill}
\centerline{\normalsize April 2004\hfill}}
\vskip1.5cm
Relation between bottom-quark $\mathrm{\overline{MS}}$ Yukawa coupling and
pole mass }

\author{Bernd A. Kniehl, Jan H. Piclum, Matthias Steinhauser
\\[.5em]
{\normalsize II. Institut f\"ur Theoretische Physik, Universit\"at Hamburg,}\\ 
{\normalsize Luruper Chaussee 149, 22761 Hamburg, Germany}
}
\date{}
\maketitle

\begin{abstract}
We calculate the $\mathcal{O}(\alpha\alpha_s)$ corrections to the 
relationships between the $\mathrm{\overline{MS}}$ Yukawa couplings and the
pole masses of the first five quark flavours in the standard model.
We also present the corresponding relationships between the
$\mathrm{\overline{MS}}$ and pole masses, which emerge as by-products of our
main analysis.
The occurring self-energies are evaluated using the method of asymptotic
expansion.

\medskip

\end{abstract}

\newpage


\section{Introduction}

One of the most intriguing puzzles in contemporary elementary particle physics
is related to the deciphering of the seemingly arbitrary pattern of the
fermion masses and the elements of the Cabibbo-Kobayashi-Maskawa (CKM) quark
mixing matrix of the standard model (SM) in the framework of some more
fundamental theory.
It is generally believed that grand unified theories (GUTs), possibly realized
in the context of supersymmetry, are able to provide a key to the
understanding of this fundamental problem \cite{gut}.
The crucial idea is that a judiciously chosen set of independent parameters,
appropriately evolved to the GUT scale, obey simple relationships.
Essential ingredients of this formalism include renormalization group 
equations, which determine the scale dependence of the running parameters, and
threshold relations, which relate the running parameters at the scales of the
elementary-particle physics experiments to the physical parameters, e.g.\
pole masses and CKM matrix elements, extracted from the latter.
It has become customary to define the running parameters in the modified
minimal-subtraction $\mathrm{\overline{MS}}$ scheme.

The Callan-Symanzik beta functions of the SM and its most popular extensions
are well established at one loop and beyond \cite{rge}.
As for the threshold relations of the Yukawa couplings and CKM matrix elements
the current status is as follows.
The relationships between the $\mathrm{\overline{MS}}$ and on-shell 
definitions of the CKM matrix elements are known at one loop \cite{Fantina}.
The relationships between the $\mathrm{\overline{MS}}$ and pole definitions of
the quark masses have been elaborated at one \cite{Braaten}, two
\cite{Broadhurst}, and three \cite{3lQCD} loops in quantum chromodynamics
(QCD).
These pure QCD corrections readily carry over to the relationships between the
$\mathrm{\overline{MS}}$ Yukawa couplings and the pole masses of the quarks,
which are actually relevant for GUT analyses.
However, the situation becomes more involved if electroweak quantum
corrections are taken into account.
Then, the relationships between the $\mathrm{\overline{MS}}$ definitions of the
fermion masses and the Yukawa couplings become nontrivial
\cite{Hempfling:1994ar,Willey}.
The full one-loop corrections to the relationships between the
$\mathrm{\overline{MS}}$ Yukawa couplings and pole masses of the SM fermions
were derived in Ref.~\cite{Hempfling:1994ar}.
They were found to be gauge-parameter independent and devoid of tadpole 
contributions.
However, in that reference, also a gauge-parameter-independent
$\mathrm{\overline{MS}}$ definition of fermion mass was presented, and tadpole
contributions were found to be essential ingredients for that.

In this paper, we take the next step and evaluate the mixed two-loop
corrections, involving one power of Sommerfeld's fine-structure constant
$\alpha$ and one power of the strong-coupling constant $\alpha_s$, to the
relationships between the $\mathrm{\overline{MS}}$ Yukawa couplings and the
pole masses of the first five quark flavours in the SM.
For simplicity, the calculation is performed in the limit where the third
quark generation does not mix with the first two, i.e.\ where
$V_{ub}=V_{cb}=V_{td}=V_{ts}=0$, which is approximately realized in nature
\cite{numeric}.
As in Ref.~\cite{Hempfling:1994ar}, we pay special attention to the tadpole
contributions.
Again, they are found to cancel in the relationships between the
$\mathrm{\overline{MS}}$ Yukawa couplings and the pole masses, while they are
indispensable to render the $\mathrm{\overline{MS}}$ definitions of quark mass
independent of the gauge parameters. 

As for the top quark, the $\mathcal{O}(\alpha\alpha_s)$ correction to the
relationship between the $\mathrm{\overline{MS}}$ and pole masses was recently
evaluated in Ref.~\cite{Jegerlehner} retaining the full mass dependence.
Also there, the tadpole contribution was found to be necessary in order to get
rid of the gauge-parameter dependence.
An alternative definition of $\mathrm{\overline{MS}}$ top-quark mass, without
inclusion of the tadpole contribution, was proposed in Ref.~\cite{Faisst}.
Since Ref.~\cite{Faisst} is dealing with leading heavy-top-quark effects, the
gauge-parameter dependence does not yet show up.
However, we caution the reader that it will once subleading corrections are
to be included in this formalism.
Then, the formulas that express physical observables in terms of the
so-defined $\mathrm{\overline{MS}}$ top-quark mass will explicitly depend on
gauge parameters as will the value of this mass.
Although this is not forbidden by first principles, it is cumbersome in
practice because, whenever a value of this mass is to be quoted, the
underlying choice of gauge needs to be specified as well
\cite{Hempfling:1994ar}.

This paper is organized as follows.
In Section~\ref{secrel}, we set up the theoretical framework and derive a
master formula for the relationship between the $\mathrm{\overline{MS}}$
Yukawa coupling and the pole mass of a quark through
$\mathcal{O}(\alpha\alpha_s)$.
In Section~\ref{secexp}, we briefly recall the method of asymptotic expansion
and apply it to recover the $\mathcal{O}(\alpha)$ results for the first five
quark flavours.
We then illustrate the goodness of the resulting expansion for the case of
bottom by comparison with the analytic $\mathcal{O}(\alpha)$ result of
Ref.~\cite{Hempfling:1994ar}.
The $\mathcal{O}(\alpha\alpha_s)$ results for the first five quark flavours
are presented in Section~\ref{sectwoloop}.
As a by-product of our analysis, we also find the relationships between the
$\mathrm{\overline{MS}}$ and pole masses of the first five quark flavours to
this order.
They can be found in Section~\ref{secmassrel}.
Finally, we conclude with a summary in Section~\ref{secsum}.


\section{Deriving the relations\label{secrel}}

The relationship between the $\mathrm{\overline{MS}}$ Yukawa coupling 
$\bar{h}_f(\mu)$ at renormalization scale $\mu$ and the pole mass $M_f$ of a
fermion $f$ can generically be written as
\begin{equation}
\bar{h}_f(\mu)=2^{3/4} G_F^{1/2} M_f \left[ 1 + \delta_f(\mu) \right],
\label{relren}
\end{equation}
where $G_F$ is Fermi's constant.
Here, we compute the corrections $\delta_f(\mu)$ for the first five quark
flavours to $\mathcal{O}(\alpha)$, $\mathcal{O}(\alpha_s)$, and
$\mathcal{O}(\alpha\alpha_s)$.

In the SM, the fermion masses are generated from the Yukawa interactions
through the Higgs mechanism of spontaneous electroweak symmetry breaking.
In terms of bare parameters, which are henceforth labelled by the subscript
zero, this is manifested through the identity
\begin{equation}
m_{f,0}=\frac{v_0}{\sqrt{2}}h_{f,0},
\label{relbare}
\end{equation}
where $v$ is the vacuum expectation value of the Higgs field.
Next, we relate the bare quantities in Eq.~(\ref{relbare}) with their
renormalized counterparts in Eq.~(\ref{relren}).

\begin{figure}[tb]
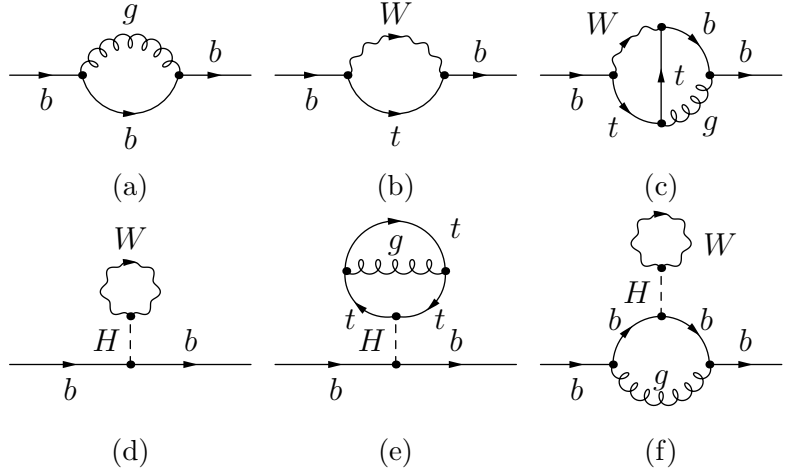

\begin{center}
\unitlength=1bp%
\begin{feynartspicture}(400,200)(3,2)

\FADiagram{(a)}
\FAProp(0.,10.)(6.,10.)(0.,){/Straight}{1}
\FALabel(3.,8.93)[t]{$b$}
\FAProp(20.,10.)(14.,10.)(0.,){/Straight}{-1}
\FALabel(17.,11.07)[b]{$b$}
\FAProp(6.,10.)(14.,10.)(0.8,){/Straight}{1}
\FALabel(10.,5.73)[t]{$b$}
\FAProp(6.,10.)(14.,10.)(-0.8,){/Cycles}{0}
\FALabel(10.,14.27)[b]{$g$}
\FAVert(6.,10.){0}
\FAVert(14.,10.){0}

\FADiagram{(b)}
\FAProp(0.,10.)(6.,10.)(0.,){/Straight}{1}
\FALabel(3.,8.93)[t]{$b$}
\FAProp(20.,10.)(14.,10.)(0.,){/Straight}{-1}
\FALabel(17.,11.07)[b]{$b$}
\FAProp(6.,10.)(14.,10.)(0.8,){/Straight}{1}
\FALabel(10.,5.73)[t]{$t$}
\FAProp(6.,10.)(14.,10.)(-0.8,){/Sine}{1}
\FALabel(10.,14.27)[b]{$W$}
\FAVert(6.,10.){0}
\FAVert(14.,10.){0}

\FADiagram{(c)}
\FAProp(0.,10.)(6.,10.)(0.,){/Straight}{1}
\FALabel(3.,8.93)[t]{$b$}
\FAProp(20.,10.)(14.,10.)(0.,){/Straight}{-1}
\FALabel(17.,11.07)[b]{$b$}
\FAProp(10.,6.)(6.,10.)(-0.434885,){/Straight}{-1}
\FALabel(6.51421,6.51421)[tr]{$t$}
\FAProp(10.,6.)(14.,10.)(0.412689,){/Cycles}{0}
\FALabel(13.4414,6.55861)[tl]{$g$}
\FAProp(10.,14.)(10.,6.)(0.,){/Straight}{-1}
\FALabel(11.07,10.)[l]{$t$}
\FAProp(10.,14.)(6.,10.)(0.425735,){/Sine}{-1}
\FALabel(6.53252,13.4675)[br]{$W$}
\FAProp(10.,14.)(14.,10.)(-0.412689,){/Straight}{1}
\FALabel(13.4414,13.4414)[bl]{$b$}
\FAVert(10.,14.){0}
\FAVert(10.,6.){0}
\FAVert(6.,10.){0}
\FAVert(14.,10.){0}

\FADiagram{(d)}
\FAProp(0.,8.)(10.,8.)(0.,){/Straight}{1}
\FALabel(5.,6.93)[t]{$b$}
\FAProp(20.,8.)(10.,8.)(0.,){/Straight}{-1}
\FALabel(15.,9.07)[b]{$b$}
\FAProp(10.,8.)(10.,12.)(0.,){/ScalarDash}{0}
\FALabel(9.18,10.)[r]{$H$}
\FAProp(10.,12.)(10.,12.)(10.,16.5){/Sine}{-1}
\FALabel(10.,17.57)[b]{$W$}
\FAVert(10.,8.){0}
\FAVert(10.,12.){0}

\FADiagram{(e)}
\FAProp(0.,8.)(10.,8.)(0.,){/Straight}{1}
\FALabel(5.,6.93)[t]{$b$}
\FAProp(20.,8.)(10.,8.)(0.,){/Straight}{-1}
\FALabel(15.,9.07)[b]{$b$}
\FAProp(10.,8.)(10.,12.)(0.,){/ScalarDash}{0}
\FALabel(9.18,10.)[r]{$H$}
\FAProp(5.8,15.75)(14.1,15.75)(-1.,){/Straight}{1}
\FALabel(14.5,18.5)[lb]{$t$}
\FAProp(10.,12.)(5.8,15.75)(-0.5,){/Straight}{1}
\FALabel(6.7,11.8)[r]{$t$}
\FAProp(10.,12.)(14.1,15.75)(0.5,){/Straight}{-1}
\FALabel(13.2,11.8)[l]{$t$}
\FAProp(5.8,15.75)(14.1,15.75)(0.,){/Cycles}{0}
\FALabel(10.,17.07)[b]{$g$}
\FAVert(10.,8.){0}
\FAVert(5.9,15.75){0}
\FAVert(14.1,15.75){0}
\FAVert(10.,12.){0}

\FADiagram{(f)}
\FAProp(0.,8.)(6.,8.)(0.,){/Straight}{1}
\FALabel(3.,6.93)[t]{$b$}
\FAProp(20.,8.)(14.,8.)(0.,){/Straight}{-1}
\FALabel(17.,9.07)[b]{$b$}
\FAProp(10.,12.)(10.,16.)(0.,){/ScalarDash}{0}
\FALabel(9.18,14.)[r]{$H$}
\FAProp(10.,16.)(10.,16.)(10.,20.5){/Sine}{-1}
\FALabel(13.5,18.)[l]{$W$}
\FAProp(10.,12.)(6.,8.)(0.4,){/Straight}{-1}
\FALabel(6.7,11.8)[r]{$b$}
\FAProp(10.,12.)(14.,8.)(-0.4,){/Straight}{1}
\FALabel(13.2,11.8)[l]{$b$}
\FAProp(6.,8.)(14.,8.)(0.8,){/Cycles}{0}
\FALabel(10.,7.5)[t]{$g$}
\FAVert(10.,12.){0}
\FAVert(10.,16.){0}
\FAVert(6.,8.){0}
\FAVert(14.,8.){0}

\end{feynartspicture}
\caption{\label{fig::self}Sample diagrams which contribute to the bottom-quark
self-energy.}
\end{center}
\end{figure}

The pole mass is defined as the zero of the inverse propagator.
The inverse fermion propagator can be written as
\begin{equation}
i S_f(q)^{-1} = \not\! q\, -m_{f,0} + \Sigma_f(q),
\label{prop}
\end{equation}
where
\begin{equation}
  \Sigma_f(q) = \sum_l \left[ \not\! q\, 
  \Sigma_{f,V}^{(l)}\left(\frac{m_{f,0}^2}{q^2}\right)
+\not\! q\, \gamma_5 \Sigma_{f,A}^{(l)}\left(\frac{m_{f,0}^2}{q^2}\right)
  + m_{f,0} \Sigma_{f,S}^{(l)}\left(\frac{m_{f,0}^2}{q^2}\right) 
  \right] 
  \label{self}
\end{equation}
is the self-energy of fermion $f$.
In Eq.~(\ref{self}), the subscripts $V$, $A$, and $S$ denote the vector,
axial-vector, and scalar components, respectively, and the sum runs over all
contributions through the desired order.
Here, we consider the one-loop contributions of $\mathcal{O}(\alpha)$ and
$\mathcal{O}(\alpha_s)$ and the two-loop ones of
$\mathcal{O}(\alpha\alpha_s)$, which we denote by the superscripts (1) and
(2), respectively.
Some sample Feynman diagrams are depicted in Fig.~\ref{fig::self}(a)--(c).
Inserting Eq.~(\ref{self}) into Eq.~(\ref{prop}) and setting the latter to
zero leads to \cite{Broadhurst}
\begin{equation}
  M_f = m_{f,0} \left\{ 1 - \Sigma_f^{(1)}(1) -  \Sigma_f^{(2)}(1) +
    \Sigma_f^{(1)}(1) \left[ \Sigma_{f,V}^{(1)}(1) - 2 \Sigma_f^{(1)\prime}(1)
    \right]\right\}, 
  \label{mpole2mbare}
\end{equation}
which is valid through the two-loop order.
Here, the prime indicates differentiation with respect to $m_{f,0}^2/q^2$,
and the notation $\Sigma_f^{(l)} = \Sigma_{f,V}^{(l)} + \Sigma_{f,S}^{(l)}$
has been introduced.
Note that, while deriving Eq.~(\ref{mpole2mbare}), an expansion of $m_{f,0}$
about $M_f$ has been performed in order to obtain unity as the argument in the
self-energies.
Furthermore, we remark that $\Sigma_{f,A}^{(l)}$ does not appear in
Eq.~(\ref{mpole2mbare}). 
For later convenience, we also provide the inverted relation, which reads
\begin{equation}
m_{f,0} = M_f \left\{ 1 + \Sigma_f^{(1)}(1) +  \Sigma_f^{(2)}(1) +
  \Sigma_f^{(1)}(1) \left[ \Sigma_{f,S}^{(1)}(1) + 2 \Sigma_f^{(1)\prime}(1)
  \right]\right\}.
\label{mbare2mpole}
\end{equation}
By construction, the mass of fermion $f$ which appears in the self-energies
on the r.h.s.\ of Eqs.~(\ref{mpole2mbare}) and (\ref{mbare2mpole}) is the pole
mass.
These equations become manifestly gauge-parameter independent if one includes
the tadpole contributions $T_f$ in the self-energies \cite{Hempfling:1994ar},
i.e.\ if the tadpole-free self-energies $\Sigma_f^{(l)}$ in
Eqs.~(\ref{mpole2mbare}) and (\ref{mbare2mpole}) are replaced by
$\Sigma_f^{(l)}+T_f^{(l)}$.
Some sample tadpole diagrams are shown in Fig.~\ref{fig::self}(d)--(f).

The Fermi constant $G_F$ is not a basic parameter of the SM Lagrangian
density.
This means that it has to be expressed in terms of such parameters.
To do this, one calculates a process in the SM and in the Fermi theory, and
equates the two results.
This was first done in a classical paper by Sirlin \cite{Sirlin:1980nh}, who
considered the muon lifetime.
Adapting his result for corrections of $\mathcal{O}(\alpha\alpha_s)$, we have
\begin{equation}
G_F = \frac{1}{\sqrt{2} v_0^2} \left[ 1 + \frac{\Pi_{WW}^{(1)}(0) +
    \Pi_{WW}^{(2)}(0)}{M_W^2} + \frac{T_{WW}^{(1)} +
    T_{WW}^{(2)}}{M_W^2} + E \right].
\label{fermi}
\end{equation}
Here, $\Pi_{WW}(q^2)$ is the tadpole-free $W$-boson self-energy at
four-momentum $q$, and $T_{WW}$ denotes the corresponding tadpole
contribution.
The quantity $E$ contains those wave-function renormalization, vertex and box
corrections to the muon decay width which the SM introduces on top of the
Fermi model improved by quantum electrodynamics (QED). 
At $\mathcal{O}(\alpha)$ and in 't~Hooft-Feynman gauge, it may be written as
\begin{equation}
E = \frac{\alpha}{4 \pi s_w^2} \left[ \frac{4}{\epsilon} + 4
  \ln\frac{\mu^2}{M_Z^2} + \left( \frac{7}{2 s_w^2} - 6 \right) \ln c_w^2 + 6
  \right],
\end{equation}
where we have used the abbreviation $c_w^2 = 1 - s_w^2 = M_W^2/M_Z^2$.
Here and in the following, we use dimensional regularization, with 
$d=4-2\epsilon$ space-time dimensions and 't~Hooft mass scale $\mu$, adopt a
$\mathrm{\overline{MS}}$-like notation where the typical combination
$\gamma_E-\ln(4\pi)$ is suppressed, and do not display terms of
$\mathcal{O}(\epsilon)$.
There are no corrections of $\mathcal{O}(\alpha\alpha_s)$ to this quantity.
At the one-loop level, the $W$-boson self-energy can be split into a bosonic
and a fermionic part as
\begin{equation}
  \Pi_{WW}(q^2) = \Pi_{WW}^{\rm bos}(q^2) + \Pi_{WW}^{\rm fer}(q^2),
\end{equation}
and, in the 't~Hooft-Feynman gauge, we have (see, e.g.,
Ref.~\cite{Hempfling:1994ar})
\begin{eqnarray}
  \Pi_{WW}^{\rm bos}(0) &=& \frac{\alpha M_W^2}{4 \pi s_w^2}\left[ \left( -2
  +\frac{1}{c_w^2} \right) \left(\frac{1}{\epsilon} + \ln \frac{\mu^2}{M_W^2}
  \right) + \left( 2 +\frac{1}{c_w^2} -\frac{17}{4 s_w^2} \right) \ln c_w^2
 \right. \nonumber\\
&&{}-\left.\frac{17}{4} +\frac{7}{8 c_w^2} 
-\frac{M_H^2}{8 M_W^2}
-\frac{3}{4}\,\frac{M_H^2}{M_W^2 - M_H^2} \ln \frac{M_W^2}{M_H^2}\right],
  \label{eq::WWbos}\\
 \Pi_{WW}^{\rm fer}(0) &=& -\frac{\alpha N_c}{8 \pi s_w^2} \left(
  m_{t,0}^2 + m_{b,0}^2 \right) \left( \frac{1}{\epsilon} + \ln
  \frac{\mu^2}{m_{t,0}^2} +\frac{1}{2} \right),
\label{eq::WWfer}
\end{eqnarray}
where $N_c=3$ denotes the number of quark colours and we have omitted terms 
quartic in $m_{b,0}$ in Eq.~(\ref{eq::WWfer}).
Note that only $\Pi_{WW}^{\rm fer}(0)$ receives $\mathcal{O}(\alpha\alpha_s)$
corrections.

Finally, the relation between the bare and the
$\mathrm{\overline{MS}}$-renormalized Yukawa couplings is given by
\begin{eqnarray}
  h_{f,0} &=& \bar{h}_f(\mu) + \delta h_f
  \nonumber\\
  &=&  \bar{h}_f(\mu) \left(1 + \delta_{f,CT}^{(1)} + 
 \delta_{f,CT}^{(2)}\right),
  \label{yukms2bare}
\end{eqnarray}
where $\delta h_f$, $\delta_{f,CT}^{(1)}$, and $\delta_{f,CT}^{(2)}$ denote
the appropriate $\mathrm{\overline{MS}}$ counterterms.

Inserting Eqs.~(\ref{mbare2mpole}), (\ref{fermi}), and (\ref{yukms2bare}) in
Eq.~(\ref{relbare}) and comparing the result with Eq.~(\ref{relren}), where
we decompose $\delta_f(\mu)=\delta_f^{(1)}(\mu)+\delta_f^{(2)}(\mu)$, we
obtain the master equations
\begin{eqnarray}
\delta_f^{(1)}(\mu)&=&\Sigma_f^{(1)}(1)-\frac{\Pi_{WW}^{(1)}(0)}{2M_W^2}
-\frac{E}{2}-\delta_{f,CT}^{(1)},
\label{delta1}\\
\delta_f^{(2)}(\mu)&=&\Sigma_f^{(2)}(1)
-\frac{\Pi_{WW}^{(2)}(0)}{2M_W^2}
+\Sigma_f^{(1)}(1)\left[\Sigma_{f,S}^{(1)}(1)+2\Sigma_f^{(1)\prime}(1)
-\frac{\Pi_{WW}^{(1)}(0)}{2M_W^2}-\frac{E}{2}\right] 
\nonumber\\
&&{}-\delta_{f,CT}^{(2)}
-\delta_{f,CT}^{(1)}\delta_f^{(1)}(\mu).
\label{delta2}
\end{eqnarray}
Here, all self-energies are tadpole-free and it is understood that all terms 
of higher orders in the electroweak couplings are discarded.
We emphasize again that Eqs.~(\ref{mbare2mpole}) and (\ref{fermi}) have to
include tadpole contributions to ensure their gauge-parameter independence.
However, in Eqs.~(\ref{delta1}) and (\ref{delta2}), these contributions cancel
thanks to the identities
\begin{eqnarray}
\frac{T_{WW}^{(1)}}{2M_W^2}&=&T_f^{(1)},
\nonumber\\
\frac{T_{WW}^{(2)}}{2M_W^2}&=&T_f^{(2)}
+T_f^{(1)}\left[\Sigma_{f,S}^{(1)}(1)+2\Sigma_f^{(1)\prime}(1)\right],
\label{eq::tad}
\end{eqnarray}
which relate the tadpole contributions to the fermion and boson self-energies
at $\mathcal{O}(\alpha)$ and $\mathcal{O}(\alpha\alpha_s)$, respectively.
We have verified Eq.~(\ref{eq::tad}) for arbitrary values of the gauge
parameters.
Note that Eqs.~(\ref{delta1}) and (\ref{delta2}) are finite and
gauge-parameter independent.
The later follows immediately from the gauge-parameter independence of
Eqs.~(\ref{mbare2mpole}) and (\ref{fermi}).
It should also be remarked that the $\mathcal{O}(\epsilon)$ terms of the
quantities $\Pi_{WW}(0)$ and $E$ are not needed explicitly, as they drop out
in the combination contained in Eq.~(\ref{delta2}).

Through $\mathcal{O}(\alpha\alpha_s)$, $\delta_f(\mu)$ may be decomposed as
\begin{equation}
\delta_f(\mu) = \frac{\alpha}{\pi} \delta_f^{(\alpha)}(\mu) 
+ \frac{\alpha_s}{\pi} C_F \delta_f^{(\alpha_s)}(\mu) 
+ \frac{\alpha\alpha_s}{\pi^2} C_F \delta_f^{(\alpha\alpha_s)}(\mu),
\end{equation}
where $C_F = (N_c^2 - 1)/(2 N_c)$ is the eigenvalue of the Casimir operator of
the fundamental representation of SU(3)$_c$.
The well-known one-loop QCD contribution reads:
\begin{equation}
\delta_f^{(\alpha_s)}(\mu) = -1 -\frac{3}{4} \ln\frac{\mu^2}{M_f^2}.
\end{equation}
The two-loop \cite{Broadhurst} and three-loop \cite{3lQCD} QCD contributions
are also known.
The $\mathcal{O}(\alpha)$ contribution is given in
Ref.~\cite{Hempfling:1994ar}.
The evaluation of the $\mathcal{O}(\alpha\alpha_s)$ one is the main purpose of
this paper.
It is outlined in Section~\ref{sectwoloop}.
Due to the presence of many different mass scales, an exact calculation is
rather cumbersome.
Thus, we adopt the concept of asymptotic expansion, which we recall and apply
to the one-loop case in the next section.


\section{Asymptotic expansion\label{secexp}}

The method of asymptotic expansion \cite{Smirnov:pj} allows for the evaluation
of Feynman integrals which depend on various parameters of different sizes.
This is achieved by expanding the integrand in small ratios of these
parameters.
As a result, one obtains an infinite series in powers and logarithms of the
expansion parameters.

In our case, the so-called hard-mass procedure is used, where one mass $M$ is
much larger than all other masses $\{m_i\}$ and external momenta $\{q_i\}$.
In this limit, the expansion can be performed in a diagrammatic way, according
to the prescription
\begin{equation}
  \Gamma (M,m,q) \stackrel{M\to \infty}{\simeq} \sum_\gamma \Gamma/\gamma(q,m)
  \star 
  \mathcal{T}_{(q_\gamma ,m_\gamma )} \gamma(M,m_\gamma ,q_\gamma ),
  \label{eq::hmp}
\end{equation}
where $\Gamma$ is the initial diagram and the operator
$\mathcal{T}_{(q_\gamma,m_\gamma )}$ performs a Taylor expansion in all small
parameters of the subdiagram $\gamma$.\
The sum runs over all subdiagrams that contain the large mass $M$ and are
one-particle-irreducible in all connected parts after contracting all lines
with the large mass.

Let us first consider the case of bottom.
In terms of the inherent mass scales, the bottom-quark self-energy diagrams
can be divided into two classes:
\begin{enumerate}
\item those involving the exchange of charged bosons ($W,\phi$), and
\item those involving the exchange of neutral bosons ($Z,\chi,H$).
\end{enumerate}
Exploiting the fact that $M_\phi=\sqrt{\xi_W}M_W$ and
$M_\chi=\sqrt{\xi_Z}M_Z$ depend on the arbitrary gauge parameters $\xi_W$ and
$\xi_Z$, we can arrange for the masses involved in these diagrams to fulfil
the following hierarchies:
\begin{eqnarray}
M_b&\ll&M_W,M_\phi\ll M_t,
\label{hiOne}\\
M_b&\ll&M_Z,M_\chi,M_H.
\label{hiTwo}
\end{eqnarray}
Since our final results are gauge-parameter independent, we can replace
Eq.~(\ref{hiOne}) by the alternative hierarchy\footnote{%
In principle, we could also choose $M_\phi,M_\chi\ll M_b$.
In this case, however, Eq.~(\ref{eq::hmp}) could not be applied, since the
external momentum is on the bottom-quark mass shell, which requires different
rules of calculation~\cite{Smirnov:pj}.}
\begin{equation}
M_b\ll M_W\ll M_t\ll M_\phi.
\label{hiThree}
\end{equation}
This changes the intermediate results for some of the diagrams, but not the
final result for $\delta_b(\mu)$.

In the case of the light quarks $q=u,d,s,c$, the situation is much simpler.  
Since we neglect their masses $m_q$, except for the linear appearances in
front of $\Sigma_{q,S}$, all diagrams are either reduced to vacuum integrals
where the scale is given by a boson mass or to on-shell integrals which can be
adopted from the bottom-quark case.

Similarly, in the case of the fermionic part of the $W$-boson self-energy,
which has to be evaluated for vanishing external momentum, we only have the
hierarchy $M_b\ll M_t$.
As for the bosonic part of the $W$-boson self-energy, we use the exact
expression given in Eq.~(\ref{eq::WWbos}).

The calculation of the self-energy diagrams can be performed in a completely
automated way.
This is achieved by the successive use of the computer programs \texttt{QGRAF}
\cite{Nogueira:1991ex}, \texttt{q2e} \cite{Seidensticker:q2e}, \texttt{exp}
\cite{Seidensticker:exp}, and \texttt{MATAD} \cite{Steinhauser:2000ry} (for a
review, see Ref.~\cite{Harlander:1998dq}).
First, \texttt{QGRAF} is used to generate the Feynman diagrams.
Its output is then rewritten by \texttt{q2e} to be understandable by
\texttt{exp}.
This program performs an asymptotic expansion of the diagrams, if necessary,
in an iterated fashion.
Finally, the \texttt{FORM}-based \cite{Vermaseren} program \texttt{MATAD} is
used to perform the actual calculations and expansions in $\epsilon$.
For the calculation of certain diagrams, it is supplemented with the package
\texttt{ON-SHELL2} \cite{Fleischer:1999tu}. 

In a first step, we explicitly check the gauge-parameter independence of the
combination $\Sigma_f^{(1)}(1) + T_f^{(1)}$ appearing in
Eq.~(\ref{mbare2mpole}).
We refrain from listing the corresponding results, which can be extracted from
the expressions given in Section~\ref{secmassrel}, where the relation between
the $\mathrm{\overline{MS}}$ and pole masses is discussed.

Let us now move on to $\delta_b^{(\alpha)}$.
Since $E$ is only known in 't~Hooft-Feynman gauge, we have to choose this 
gauge also for $\Sigma_f^{(1)}(1)$ and $\Pi_{WW}^{(1)}(0)$.
We determine the counterterm $\delta_{f,CT}^{(1)}$ in Eq.~(\ref{yukms2bare})
by requiring that Eq.~(\ref{delta1}) be finite and thus find
\begin{eqnarray}
\delta_{b,CT}^{(1)}&=&
-\frac{\alpha_s}{\pi}C_F\frac{3}{4}\,\frac{1}{\epsilon}
+\frac{\alpha}{\pi}\left[
-\frac{3}{4} v_b^2
-\frac{3}{4} Q_b^2 -\frac{1}{s_w^2} \left( \frac{3}{64 c_w^2} +\frac{3}{16}
\right)\right.
\nonumber\\
&&{}+\left.
\frac{m_{t,0}^2}{M_W^2}\,
\frac{1}{s_w^2} \left(- \frac{3}{32} +\frac{N_c}{16} \right)
+\frac{m_{b,0}^2}{M_W^2}\,
\frac{1}{s_w^2} \left( \frac{3}{32} +\frac{N_c}{16} \right)
\right]\frac{1}{\epsilon}.
\end{eqnarray}
Inserting all the above ingredients into Eq.~(\ref{delta1}), we obtain
\begin{eqnarray}
\delta_b^{(\alpha)}(\mu)&=&
%
%
\frac{M_t^2}{M_W^2} \frac{N_c}{s_w^2}
  \Bigg[ \frac{1}{32} +\frac{1}{16} l_t \Bigg] +\frac{M_t^2}{M_W^2}
  \frac{1}{s_w^2} \Bigg[ -\frac{5}{64} -\frac{3}{32} l_t \Bigg] +
  \frac{M_H^2}{M_W^2} \frac{1}{s_w^2} \frac{1}{64} \nonumber \\
%
%
&&{}+\frac{M_H^2}{M_W^2 - M_H^2} \frac{1}{s_w^2} \frac{3}{32} l_{WH}
   +\frac{1}{s_w^2} \Bigg[ -\frac{5}{32} -\frac{3}{16} l_W \Bigg]
   + \frac{1}{s_w^2 c_w^2} \Bigg[ -\frac{11}{128} -\frac{3}{64} l_Z \Bigg]
   \nonumber \\
&&{}+\frac{1}{s_w^4} \frac{3}{32} \ln c_w^2
   + Q_b^2 \Bigg[ -1 -\frac{3}{4} l_b \Bigg]
   + v_b^2 \Bigg[ -\frac{5}{8} -\frac{3}{4} l_Z \Bigg]
%
%
   +\frac{M_W^2}{M_t^2} \frac{1}{s_w^2} \Bigg[ \frac{3}{32} +\frac{3}{32}
l_{Wt} \Bigg] \nonumber \\
%
%
&&{}+\frac{M_W^4}{M_t^4} \frac{1}{s_w^2} \Bigg[ \frac{3}{32} +\frac{3}{16}
l_{Wt} \Bigg]
%
%
 +\frac{M_W^6}{M_t^6} \frac{1}{s_w^2} \Bigg[ \frac{3}{32} +\frac{9}{32}
l_{Wt} \Bigg] \nonumber \\
%
%
&&{} +\frac{M_W^8}{M_t^8} \frac{1}{s_w^2} \Bigg[ \frac{3}{32} +\frac{3}{8}
l_{Wt} \Bigg]
%
%
 +\frac{M_W^{10}}{M_t^{10}} \frac{1}{s_w^2} \Bigg[ \frac{3}{32}
+\frac{15}{32} l_{Wt} \Bigg] \nonumber \\
%
%
&&{}+\frac{M_b^2}{M_W^2} \Bigg\{ \frac{1}{s_w^2} \Bigg[ \frac{N_c}{32} +
\frac{11}{192} +\frac{N_c}{16} l_t -\frac{1}{16} l_b +\frac{1}{32} l_t
+\frac{1}{32} l_Z +\frac{3}{32} l_H \Bigg]
 \nonumber \\
&&{} +\frac{M_W^2}{M_Z^2} v_b^2 \Bigg[ -\frac{2}{3} -l_{bZ} \Bigg]
\Bigg\} 
+\cdots.
\label{delta1l}
\end{eqnarray}
Here and in the following, ellipses stand for terms of
$\mathcal{O}(M_W^{12}/M_t^{12})$, $\mathcal{O}(M_W^2M_b^2/M_t^4)$, or
$\mathcal{O}(M_b^4/M_W^4)$.
As we will see below, they are very small and can safely be neglected.
Note that there are no terms proportional to $M_b^2/M_t^2$ in
Eq.~(\ref{delta1l}).
Here and in the following, $Q_f$ is the fractional electric charge of fermion
$f$, $I_{3,f}$ is the third component of weak isospin of its left-handed
component, $v_f = (I_f^3 - 2 s_w^2 Q_f)/(2 c_w s_w)$ and
$a_f = I_f^3/(2 c_w s_w)$ are its vector and axial-vector couplings to the
$Z$ boson, respectively, and we use the abbreviations
$l_i = \ln (\mu^2/M_i^2)$ and $l_{ij} = \ln (M_i^2/M_j^2)$,

For the light quarks $q=u, d, s, c$, we have
\begin{eqnarray}
  \delta_q^{(\alpha)}(\mu) &=& \frac{M_t^2}{M_W^2} \frac{N_c}{s_w^2}
  \Bigg[ \frac{1}{32} +\frac{1}{16} l_t \Bigg] +\frac{M_H^2}{M_W^2}
  \frac{1}{s_w^2}\frac{1}{64} +\frac{M_H^2}{M_W^2 - M_H^2} \frac{1}{s_w^2}
  \frac{3}{32} l_{WH}
\nonumber\\
&&{}+\frac{1}{s_w^2} \Bigg[ -\frac{1}{4} -\frac{3}{16} l_W \Bigg]
  +\frac{1}{s_w^2 c_w^2} \Bigg[ -\frac{11}{128}
  -\frac{3}{16} l_Z \Bigg] +\frac{1}{s_w^4} \frac{3}{32} \ln c_w^2\nonumber \\
&&{}  + Q_q^2 \Bigg[ -1 -\frac{3}{4} l_q
  \Bigg] + v_q^2 \Bigg[ -\frac{5}{8} -\frac{3}{4} l_Z \Bigg] 
  +\frac{M_b^2}{M_W^2} \frac{N_c}{s_w^2} \Bigg[ \frac{1}{32} +\frac{1}{16}
  l_t \Bigg]
+\cdots.
\label{eq::delta_light}
\end{eqnarray}

\begin{figure}[tb]
\begin{center}
\epsfig{file=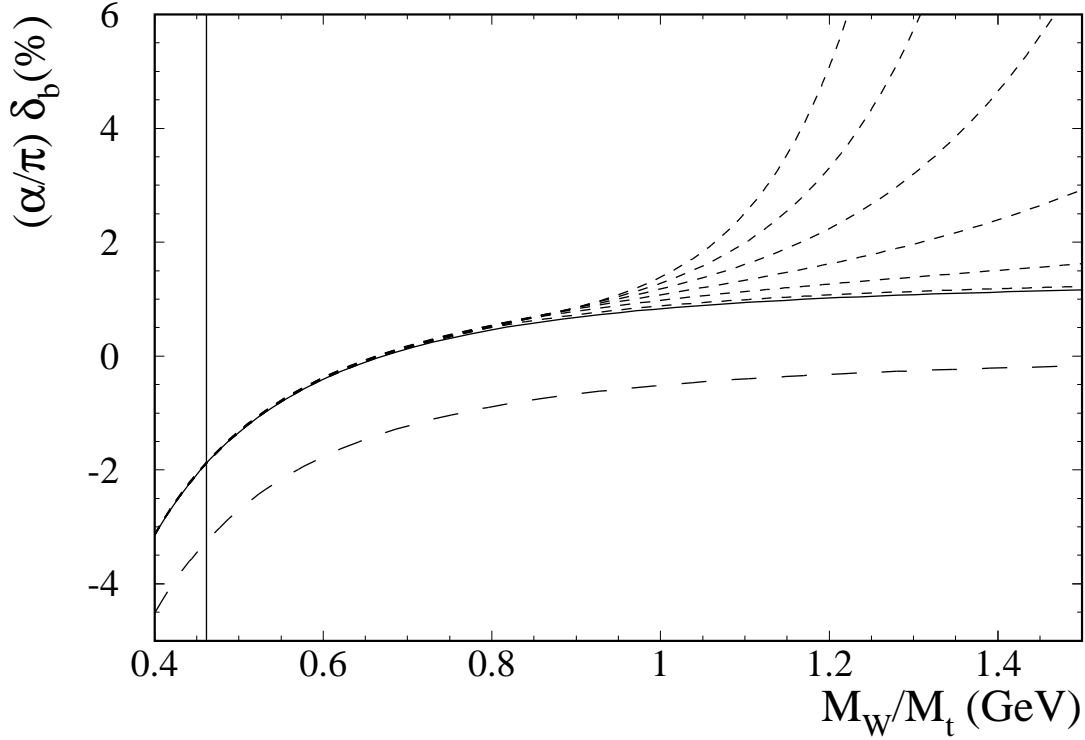,width=\textwidth}
\caption{\label{figCompExp}Comparison of our result for
$(\alpha/\pi)\delta_b^{(\alpha)}(M_b)$ with the exact result.
The latter is shown as a continuous line.
The long-dashed line indicates the leading-order contribution proportional to
$M_t^2$ and $M_H^2$ only.
The contributions given by successively adding the subleading terms are
indicated by the short-dashed lines.
The horizontal line marks the actual value of $M_W/M_t$.
Note that for this plot $M_W$ is fixed to its default value and $M_t$ is
varied.}
\end{center}
\end{figure}

Let us now discuss the quality of our approximation by comparing 
$\delta_b^{(\alpha)}(\mu)$ of Eq.~(\ref{delta1l}) with the exact result
obtained in Ref.~\cite{Hempfling:1994ar}.
As input values for our numerical analysis, we use $\alpha=1/137.035$,
$M_b=4.5$~GeV, $M_t=174.3$~GeV, $M_W=80.423$~GeV, $M_Z=91.1876$~GeV
\cite{numeric}, and $M_H=120$~GeV.
Furthermore, we choose $\mu=M_b$.
Figure~\ref{figCompExp} displays our result as well as the exact one as
functions of $M_W/M_t$.
To illustrate the convergence of the expansion, we show the leading-order
contribution separately and then successively add the subleading terms.
It is obvious from this figure that the series in $M_W^2/M_t^2$ converges very
fast.
Indeed, in the range $M_W/M_t\alt0.8$, the sum of all calculated terms can
barely be distinguished from the exact result.
The approximation including the $\mathcal{O}(M_W^{10}/M_t^{10})$ terms
actually provides an excellent approximation also for $M_W/M_t\approx1.5$,
whereas the lower-order approximations start to exhibit significant deviations
at $M_W/M_t\approx1$.

\section{Two-loop result and numerical analysis\label{sectwoloop}}

In order to obtain $\delta_f(\mu)$ to $\mathcal{O}(\alpha\alpha_s)$, one
has to evaluate the respective two-loop corrections to $\Sigma_f(q)$ and
$\Pi_{WW}(0)$, as can be seen from Eq.~(\ref{delta2}).
The result for the latter is well-known (see, e.g.,
Ref.~\cite{Djouadi:1993ss}).
Including terms quadratic in $m_b$, one has
\begin{eqnarray}
\Pi_{WW}^{(2)}(0)&=&\frac{\alpha\alpha_s}{\pi^2}\,\frac{C_F N_c}{s_w^2}
\left\{-\left( m_{t,0}^2 + m_{b,0}^2 \right) \left[ \frac{3}{32}\,
    \frac{1}{\epsilon^2} + \left(-\frac{1}{64} + \frac{3}{16} \ln
      \frac{\mu^2}{m_{t,0}^2} \right) \frac{1}{\epsilon}
\right.\right.\nonumber\\
&&{}-\left.\frac{1}{32}\ln\frac{\mu^2}{m_{t,0}^2}
+\frac{3}{16}\ln^2\frac{\mu^2}{m_{t,0}^2}\right]
-m_{t,0}^2\left[\frac{7}{128}+\frac{\zeta(2)}{32}\right] 
\nonumber\\
&&{}-\left.m_{b,0}^2\left[\frac{15}{128}+\frac{5}{32}\zeta(2)\right]\right\}
+\cdots,
\end{eqnarray}
where $\zeta$ denotes Riemann's zeta function, with the value
$\zeta(2)=\pi^2/6$.
We again refrain from listing explicitly our results for the fermion
self-energies, but provide the results for $\delta_f^{(2)}(\mu)$, from which
the expressions for $\Sigma_f^{(2)}(1)$ can be obtained if the result for the
counterterm $\delta_{f,CT}^{(2)}$ introduced in Eq.~(\ref{yukms2bare}) is
given.
Similarly to the one-loop case discussed in the previous section, 
$\delta_{f,CT}^{(2)}$ is determined by requiring that Eq.~(\ref{delta2}) be
finite upon renormalization of the top- and bottom-quark masses, and we find
\begin{eqnarray}
\delta_{b,CT}^{(2)}&=&
\frac{\alpha \alpha_s}{\pi^2} C_F\left\{
\left[\frac{9}{16}\left(Q_b^2+ v_b^2\right)
+\frac{1}{s_w^2} \left( \frac{9}{256 c_w^2} +\frac{9}{64} \right)\right]
\frac{1}{\epsilon^2}\right.
\nonumber\\
&&{}+\left[
-\frac{3}{32}\left(Q_b^2+v_b^2\right)+\frac{21}{32} a_b^2
+\frac{9}{128 s_w^2}
+\frac{m_{t,0}^2}{M_W^2}\,
\frac{1}{s_w^2} \left( -\frac{3}{32}+\frac{5}{128} N_c \right)
\right.\nonumber\\
&&{}+\left.\left.\frac{m_{b,0}^2}{M_W^2}\,
\frac{1}{s_w^2} \left( \frac{3}{32} +\frac{5}{128} N_c \right)
\right]\frac{1}{\epsilon}\right\}.
\label{deltabansatz}
\end{eqnarray}
Note that there are no poles proportional to $m_{t,0}^4$ or $m_{b,0}^4$,
although such terms do appear in certain tadpole diagrams.
However, the tadpole contributions cancel in the computation of
$\delta_b(\mu)$, as we have explained in Section~\ref{secrel}.
As an additional check, we can exploit the fact that the non-local logarithmic
terms have to cancel as well.

Finally, the $\mathcal{O}(\alpha\alpha_s)$ correction to the relationship
between the $\mathrm{\overline{MS}}$ Yukawa coupling and the pole mass of the
bottom quark turns out to be
\begin{eqnarray}
\delta_b^{(\alpha\alpha_s)}(\mu)&=&
%
%
\frac{M_t^2}{M_W^2} \frac{N_c}{s_w^2} \Bigg[ \frac{21}{256} -
  \frac{\zeta (2)}{32} - \frac{3}{128} l_b -
  \frac{7}{64} l_t -\frac{3}{64} l_b l_t - \frac{3}{64} l_t^2 \Bigg] \nonumber\\
&&{}+\frac{M_t^2}{M_W^2} \frac{1}{s_w^2} \Bigg[ -\frac{13}{64} + \frac{3}{16}
  \zeta (2) +\frac{15}{256} l_b +\frac{3}{32} l_t +\frac{9}{128} l_b l_t
  +\frac{9}{128} l_t^2 \Bigg] \nonumber\\
&&{}+\frac{M_H^2}{M_W^2} \frac{1}{s_w^2} \Bigg[ -\frac{1}{64} -\frac{3}{256}
l_b \Bigg] 
%
%
+ \frac{M_H^2}{M_W^2 - M_H^2} \frac{l_{WH}}{s_w^2} \Bigg[ -\frac{3}{32}
  -\frac{9}{128} l_b \Bigg] \nonumber\\
&&{}+\frac{1}{s_w^2} \Bigg[ -\frac{71}{256} +\frac{3}{16} \zeta (2)
+\frac{15}{128} l_b +\frac{9}{64} l_t +\frac{3}{16} l_W +\frac{9}{64} l_b
l_W \Bigg] \nonumber \\
&&{}+\frac{1}{s_w^2 c_w^2} \Bigg[ \frac{135}{1024} +\frac{33}{512} l_b
+\frac{33}{256} l_Z +\frac{9}{256} l_b l_Z \Bigg] +\frac{\ln c_w^2}{s_w^4}
\Bigg[ -\frac{3}{32} -\frac{9}{128} l_b \Bigg] \nonumber \\
&&{}+ Q_b^2 \Bigg[ \frac{7}{64} +6\zeta (2) \ln 2 -\frac{15}{4} \zeta (2)
-\frac{3}{2} \zeta (3)
+\frac{21}{16} l_b +\frac{9}{16} l_b^2 \Bigg] \nonumber\\
&&{}+v_b^2 \Bigg[ \frac{23}{64} +\frac{15}{32} l_b
+\frac{9}{16} l_Z +\frac{9}{16} l_b l_Z \Bigg] \nonumber\\
%
%
&&{}+\frac{M_W^2}{M_t^2} \frac{1}{s_w^2} \Bigg[ -\frac{81}{128} +\frac{21}{64}
\zeta (2) -\frac{9}{128} l_b -\frac{3}{128} l_{Wt} -\frac{9}{128} l_b l_{Wt}
\Bigg] \nonumber\\
%
%
&&{}+\frac{M_W^4}{M_t^4} \frac{1}{s_w^2} \Bigg[ -\frac{305}{384} +\frac{15}{32}
\zeta (2) -\frac{9}{128} l_b -\frac{5}{64} l_{Wt} -\frac{9}{64} l_b l_{Wt}
\Bigg] \nonumber\\
%
%
&&{}+\frac{M_W^6}{M_t^6} \frac{1}{s_w^2} \Bigg[ -\frac{377}{384} +\frac{39}{64}
\zeta (2) -\frac{9}{128} l_b -\frac{5}{64} l_{Wt} -\frac{27}{128} l_b l_{Wt}
\Bigg] \nonumber\\
%
%
&&{}+\frac{M_W^8}{M_t^8} \frac{1}{s_w^2} \Bigg[ -\frac{22669}{19200}
+\frac{3}{4} \zeta (2) -\frac{9}{128} l_b
-\frac{13}{320} l_{Wt} -\frac{9}{32} l_b l_{Wt} \Bigg] \nonumber\\
%
%
&&{}+\frac{M_W^{10}}{M_t^{10}} \frac{1}{s_w^2} \Bigg[ -\frac{106451}{76800}
+\frac{57}{64} \zeta (2) -\frac{9}{128} l_b
+\frac{33}{1280} l_{Wt} -\frac{45}{128} l_b l_{Wt} \Bigg] \nonumber\\
%
%
&&{}+\frac{M_b^2}{M_W^2} \Bigg\{ \frac{N_c}{s_w^2} \Bigg[ \frac{29}{256}
+\frac{\zeta (2)}{32} -\frac{9}{128} l_b -\frac{1}{16} l_t -\frac{9}{64} l_b
l_t +\frac{3}{64} l_t^2 \Bigg] \nonumber\\
&&{}+\frac{1}{s_w^2} \Bigg[ \frac{77}{216} +\frac{\zeta (2)}{24}
-\frac{229}{2304} l_b -\frac{89}{1152} l_t +\frac{19}{1152} l_Z +\frac{3}{128}
l_H \nonumber\\
&&{}-\frac{9}{128} l_b l_t +\frac{17}{384} l_b l_Z -\frac{27}{128} l_b l_H
+\frac{7}{192} l_b^2 +\frac{3}{128} l_t^2 -\frac{13}{384} l_Z^2
 +\frac{9}{128} l_H^2 \Bigg]\nonumber\\
&&{}+\frac{M_W^2}{M_Z^2} v_b^2 \Bigg[
  -\frac{25}{18} +\zeta (2) -\frac{8}{9} l_b  +\frac{25}{18} l_Z
  +\frac{13}{12} l_b l_Z -\frac{11}{12} l_b^2 -\frac{1}{6} l_Z^2 \Bigg] \nonumber\\
&&{}+\frac{M_W^2}{M_t^2} \frac{1}{s_w^2} \Bigg[
-\frac{13}{96} +\frac{\zeta (2)}{16} +\frac{5}{96} l_{bt} \Bigg] 
\Bigg\}+ \cdots.
\label{eq::deltab}
\end{eqnarray}
We should mention that the QED portion of the $\mathcal{O}(\alpha\alpha_s)$ 
contribution can also be inferred from the two-loop $\mathcal{O}(\alpha_s^2)$ 
one \cite{Broadhurst} by replacing $\alpha_s^2 C_F^2$ with
$2 \alpha_s C_F \alpha Q_f^2$ and setting all the other colour structures to
zero.
In order to obtain the result for $\delta_b^{(\alpha\alpha_s)}(\mu)$ in
Eq.~(\ref{eq::deltab}), we have used the tadpole-free bottom-quark self-energy
and adopted the 't~Hooft-Feynman gauge, the one in which $E$ is given.
The gauge-parameter independence of the full self-energy is discussed in the
next section.

In the case of the light quarks $q=u,d,s,c$, we obtain the compact expression
\begin{eqnarray}
\delta_q^{(\alpha\alpha_s)}(\mu)&=&
\frac{M_t^2}{M_W^2} \frac{N_c}{s_w^2}
  \Bigg[ \frac{21}{256} - \frac{\zeta (2)}{32} - \frac{3}{128} l_q -
  \frac{7}{64} l_t -\frac{3}{64} l_q l_t - \frac{3}{64} l_t^2 \Bigg]
  \nonumber\\
&&{}+\frac{M_H^2}{M_W^2} \frac{1}{s_w^2} \Bigg[ -\frac{1}{64} -\frac{3}{256}
  l_q \Bigg] + \frac{M_H^2}{M_W^2 - M_H^2} \frac{l_{WH}}{s_w^2} \Bigg[ -\frac{3}{32}
  -\frac{9}{128} l_b \Bigg] \nonumber\\
&&{}+\frac{1}{s_w^2} \Bigg[ \frac{79}{256} +\frac{3}{16} l_q +\frac{21}{64} l_W
 +\frac{9}{64} l_q l_W \Bigg] \nonumber \\
&&{}+\frac{1}{s_w^2 c_w^2} \Bigg[ \frac{135}{1024} +\frac{33}{512} l_q
+\frac{33}{256} l_Z +\frac{9}{256} l_q l_Z \Bigg]
  +\frac{\ln c_w^2}{s_w^4} \Bigg[ -\frac{3}{32} -\frac{9}{128} l_q \Bigg]
\nonumber \\
&&{}+ Q_q^2 \Bigg[ \frac{7}{64} +6\zeta (2) \ln 2 -\frac{15}{4} \zeta (2)
  -\frac{3}{2} \zeta (3) +\frac{21}{16} l_q +\frac{9}{16} l_q^2 \Bigg]
\nonumber\\
&&{}+v_q^2 \Bigg[ \frac{23}{64} +\frac{15}{32} l_q +\frac{9}{16} l_Z
  +\frac{9}{16} l_q l_Z \Bigg] \nonumber\\
&&{} +\frac{M_b^2}{M_W^2} \frac{N_c}{s_w^2} \Bigg[ \frac{29}{256} +\frac{\zeta
  (2)}{32} -\frac{3}{128} l_q -\frac{3}{64} l_b -\frac{1}{16} l_t
-\frac{3}{64} l_q l_t -\frac{3}{32} l_b l_t +\frac{3}{64} l_t^2 \Bigg]
+\cdots.
\nonumber\\
&&
\end{eqnarray}

\begin{figure}[tb]
\begin{center}
\epsfig{file=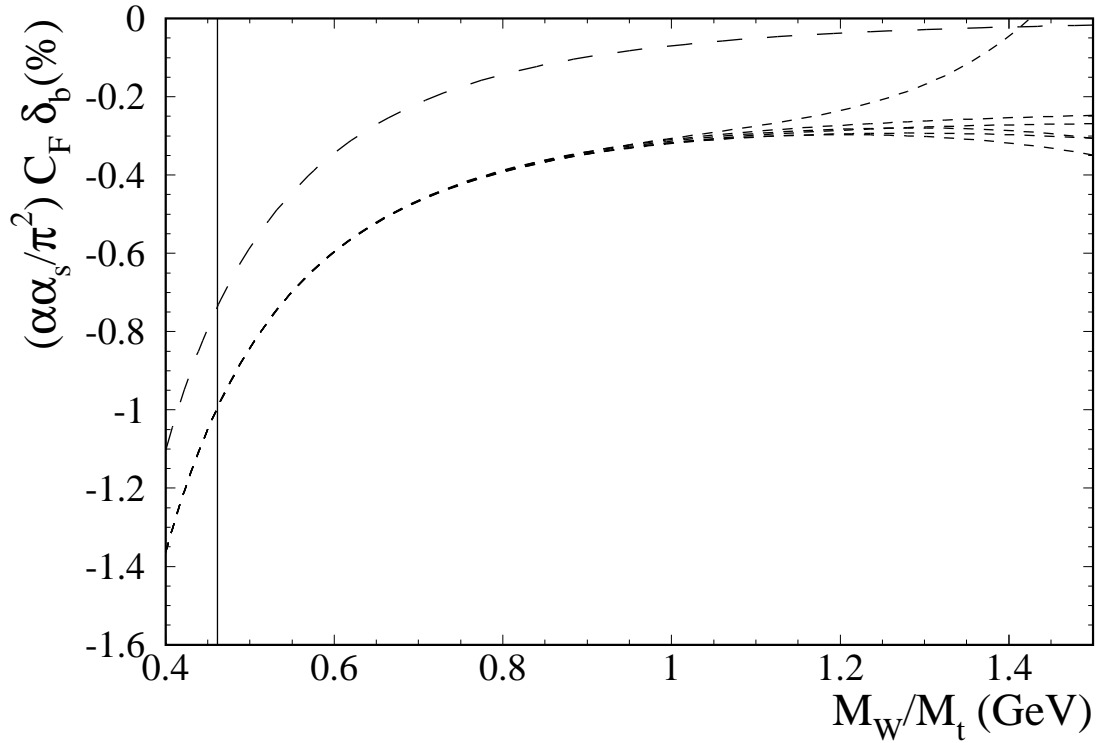,width=\textwidth}
\caption{\label{figRes2lconv}
$(\alpha\alpha_s/\pi^2)C_F\delta_b^{(\alpha\alpha_s)}(M_b)$ as a function of
$M_W/M_t$ for $M_H=120$~GeV. 
The long-dashed line indicates the leading-order contribution only.
The contributions given by successively adding the subleading terms are
indicated by the short-dashed lines.
The horizontal line marks the actual value of $M_W/M_t$.
Note that for this plot $M_W$ is fixed to its default value and $M_t$ is
varied.}
\end{center}
\end{figure}

In order to demonstrate the good convergence properties of our result, we show 
in Fig.~\ref{figRes2lconv} the $\mathcal{O}(\alpha\alpha_s)$ contribution to
$\delta_b(M_b)$ as a function of $M_W/M_t$ for $M_H=120$~GeV.
In addition to the parameters specified after Eq.~(\ref{eq::delta_light}), we
use $\alpha_s(M_b)=0.1905$ appropriate for six active quark flavours, which we
evaluate from the present world average $\alpha_s^{(5)}(M_Z)=0.1172$
\cite{numeric} using the program \texttt{RunDec} \cite{Chetyrkin:2000yt}.
Again, we observe an excellent convergence up to $M_W/M_t\approx0.8$, which is
more than sufficient for our purpose.
Since it is not possible to distinguish between the different lines in the
physically interesting region in Fig.~\ref{figRes2lconv}, we give explicit
numbers for the subleading terms at $M_W/M_t\approx0.46$.
For this purpose, we write the series as
\begin{equation}
\delta_b^{(\alpha\alpha_s)}(M_b) = C_{l,1} \frac{M_t^2}{M_W^2} + C_{l,2}
\frac{M_H^2}{M_W^2} +\sum_{k=0}^\infty C_k \left( \frac{M_W^2}{M_t^2}
\right)^k
\end{equation}
and thus obtain
\begin{eqnarray}
\lefteqn{\sum_{n=0}^5 \frac{C_n \left(M_W^2/M_t^2\right)^n}
{\sum\limits_{m=0}^5 C_m \left(M_W^2/M_t^2\right)^m}}
  \nonumber \\
&& = 0.997925 + 0.004058 - 0.001451 - 0.000448 - 0.000078 - 0.000006.
\end{eqnarray}
This demonstrates that, aside from the leading-order terms, the
$\mathcal{O}(M_W^0/M_t^0)$ term yields the largest contribution by far, while
all other subleading terms are rather small.
Furthermore, we wish to mention that the inclusion of the corrections
quadratic in $M_b$ have no visible effect in Fig.~\ref{figRes2lconv}.

\begin{figure}[tb]
\begin{center}
\epsfig{file=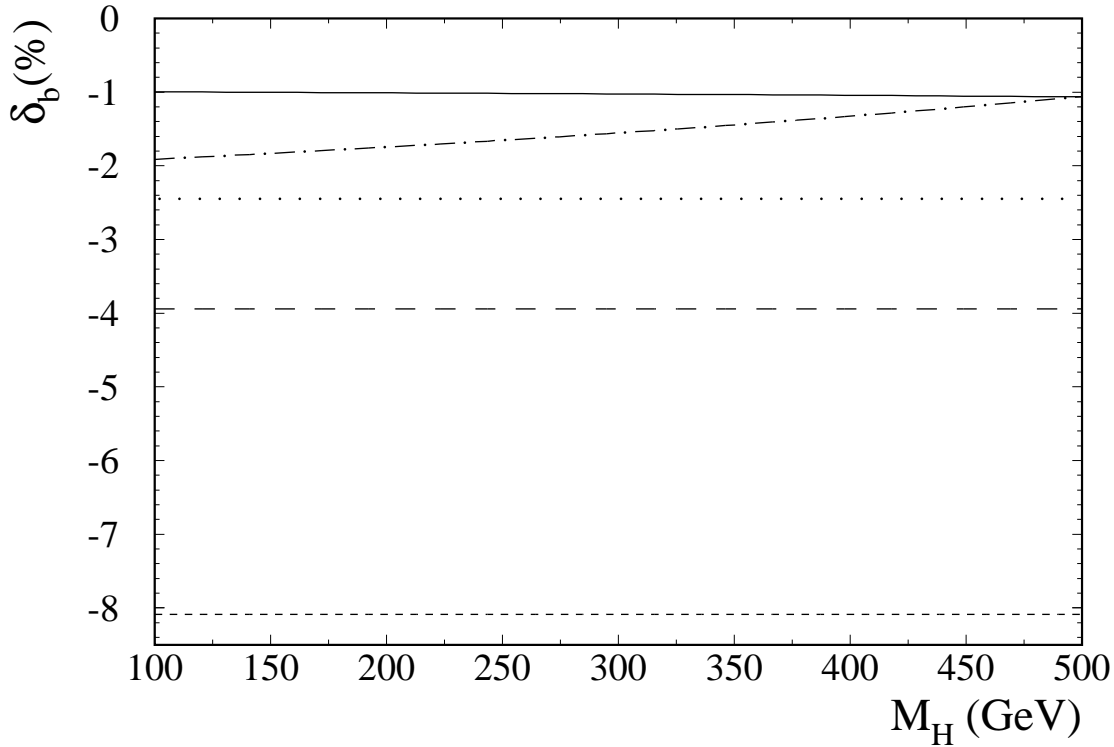,width=\textwidth}
\caption{\label{figRes2lMH}$\delta_b(M_b)$ as a function of $M_H$.
The dash-dotted and full lines indicate the $\mathcal{O}(\alpha)$ and
$\mathcal{O}(\alpha\alpha_s)$ contributions, respectively.
For comparison, also the pure QCD contributions of $\mathcal{O}(\alpha_s)$
(short-dashed line), $\mathcal{O}(\alpha_s^2)$ (long-dashed line), and
$\mathcal{O}(\alpha_s^3)$ (dotted line) are shown.
The latter are, of course, independent of $M_H$.}
\end{center}
\end{figure}

\begin{figure}[tb]
\begin{center}
\epsfig{file=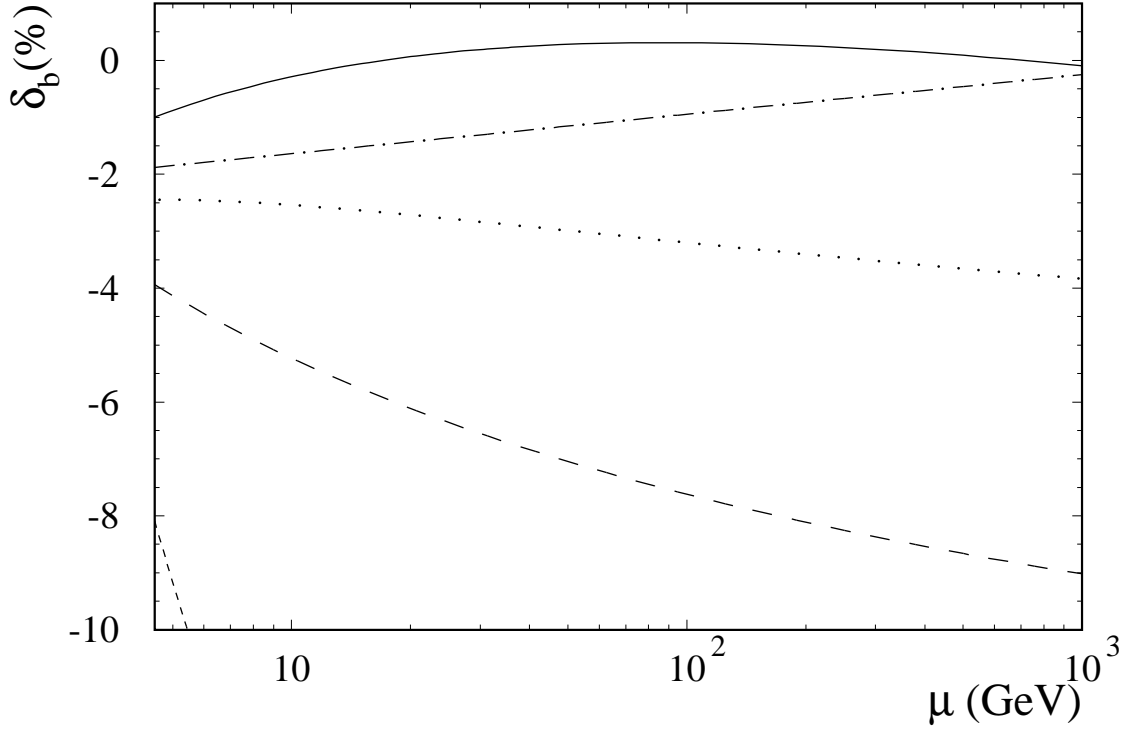,width=\textwidth}
\caption{\label{figRes2lmu}$\delta_b(\mu)$ as a function of $\mu$ for
$M_H=120$~GeV.
The same coding as in Fig.~\ref{figRes2lMH} is adopted.}
\end{center}
\end{figure}

Our final numerical results are presented in Figs.~\ref{figRes2lMH} and
\ref{figRes2lmu}.
Figure~\ref{figRes2lMH} shows the $\mathcal{O}(\alpha)$ and
$\mathcal{O}(\alpha\alpha_s)$ contributions to $\delta_b(M_b)$ as functions of
$M_H$.
For comparison, also the pure QCD ones of $\mathcal{O}(\alpha_s)$,
$\mathcal{O}(\alpha_s^2)$, and $\mathcal{O}(\alpha_s^3)$ are plotted.
We observe that the $\mathcal{O}(\alpha\alpha_s)$ contribution exhibits a
rather weak dependence on $M_H$.
Futhermore, for $M_H\approx 500$~GeV, it becomes comparable in size to the
$\mathcal{O}(\alpha)$ contribution, whose magnitude decreases with increasing 
value of $M_H$.
From Fig.~\ref{figRes2lMH}, we also see that our new
$\mathcal{O}(\alpha\alpha_s)$ contribution is of the same order of magnitude
as the $\mathcal{O}(\alpha_s^3)$ one.

Figure~\ref{figRes2lmu} shows $\delta_b(\mu)$ for $M_H=120$~GeV as a function
of $\mu$ in the range $M_b<\mu<1000$~GeV.
We observe that the $\mathcal{O}(\alpha\alpha_s)$ contribution takes positive
values in the range $20\alt\mu\alt600$~GeV.
It is comparable in size to the $\mathcal{O}(\alpha_s^3)$ one only in
the lower $\mu$ range.


\section{Relationship between $\mathbf{\overline{MS}}$ and pole masses
\label{secmassrel}} 

As a by-product of our calculation, we obtain the
$\mathcal{O}(\alpha\alpha_s)$ correction to the relationship between the
$\mathrm{\overline{MS}}$ and pole definitions of mass for the first five quark
flavours.
As discussed in Section~\ref{secrel}, it is necessary to include the tadpole
contributions in order to obtain gauge-parameter-independent results.
This has been checked explicitly by considering the hierarchies discussed in
Eqs.~(\ref{hiOne}), (\ref{hiTwo}), and (\ref{hiThree}).

The starting point for the derivation of the $\mathrm{\overline{MS}}$ to
on-shell relationship is Eq.~(\ref{mbare2mpole}), where we replace the bare
quark mass by the $\mathrm{\overline{MS}}$ one.
The corresponding relation is obtained in analogy to Eq.~(\ref{deltabansatz}),
with the only difference that we also have to allow for terms quartic in
$m_{b,0}$ and $m_{t,0}$ because of the tadpole contribution.
By requiring finiteness of the resulting relation, we obtain
\begin{eqnarray}
\frac{m_{b,0}}{\overline{m}_b}&=&
 1 -\frac{\alpha_s}{\pi} C_F \frac{3}{4}
  \frac{1}{\epsilon} + \frac{\alpha}{\pi} \left[ -\frac{3}{4} \left(Q_b^2
 +v_b^2-a_b^2 \right) -\frac{3}{16 c_w^2 s_w^2} \frac{M_Z^2}{M_H^2} -
  \frac{3}{32 s_w^2} \frac{M_H^2}{M_W^2}
\right.\nonumber\\
&&{}-\left. \frac{3}{8 s_w^2} \frac{M_W^2}{M_H^2}
- \frac{3}{32 s_w^2} \frac{\overline{m}_t^2}{M_W^2} +\frac{N_c}{4 s_w^2}
  \frac{\overline{m}_t^4}{M_W^2 M_H^2} +\frac{3}{32 s_w^2}
\frac{\overline{m}_b^2}{M_W^2} \right] \frac{1}{\epsilon}
\nonumber\\
&&{}+\frac{\alpha\alpha_s}{\pi^2} C_F \Bigg\{ \left[
  \frac{9}{16} \left(Q_b^2+v_b^2-a_b^2\right) +\frac{9}{64 c_w^2 s_w^2}
  \frac{M_Z^2}{M_H^2} +\frac{9}{128 s_w^2} \frac{M_H^2}{M_W^2}
\right.\nonumber\\
&&{}+\left.
\frac{9}{32 s_w^2} \frac{M_W^2}{M_H^2} + \frac{9}{64 s_w^2}
 \frac{\overline{m}_t^2}{M_W^2} -\frac{9 N_c}{16 s_w^2}
  \frac{\overline{m}_t^4}{M_W^2 M_H^2} -\frac{9}{64 s_w^2}
  \frac{\overline{m}_b^2}{M_W^2} \right] \frac{1}{\epsilon^2}
\nonumber\\
&&{}+\left[
  -\frac{3}{32} \left(Q_b^2+v_b^2\right) +\frac{21}{32} a_b^2
  +\frac{9}{128s_w^2}
\right.\nonumber\\
&&{}-\left.\left.
\frac{3}{32 s_w^2} \frac{\overline{m}_t^2}{M_W^2} +\frac{N_c}{8 s_w^2}
  \frac{\overline{m}_t^4}{M_W^2 M_H^2} +\frac{3}{32 s_w^2}
  \frac{\overline{m}_b^2}{M_W^2} \right] \frac{1}{\epsilon} \right\}
+\cdots. 
\end{eqnarray}
If the tadpole contributions were omitted here, then there would be no terms
quartic in $\overline{m}_b$ and $\overline{m}_t$ and no terms involving $M_H$.
In turn, gauge-parameter dependence would appear in the nonleading terms.

The relationship between the $\mathrm{\overline{MS}}$ and pole masses of the
bottom quark through $\mathcal{O}(\alpha\alpha_s)$ reads:
\begin{eqnarray}
\frac{\overline{m}_b(\mu)}{M_b} &=& 1
+\frac{\alpha_s}{\pi} C_F 
\Bigg\{ -1 -\frac{3}{4} l_b \Bigg\}
%
%
+\frac{\alpha}{\pi}
%
%
\Bigg\{ \frac{M_t^4}{M_W^2 M_H^2}\frac{N_c}{s_w^2}
  \Bigg[ \frac{1}{4} +\frac{1}{4} l_t \Bigg] +\frac{M_t^2}{M_W^2}
  \frac{1}{s_w^2} \Bigg[ -\frac{5}{64} -\frac{3}{32} l_t \Bigg]
  \nonumber \\ 
&&{}
  +\frac{M_H^2}{M_W^2} \frac{1}{s_w^2} \Bigg[ -\frac{3}{32} -\frac{3}{32} l_H \Bigg]
%
%
%
 + Q_b^2 \Bigg[ -1 -\frac{3}{4} l_b \Bigg] + v_b^2 \Bigg[ -\frac{5}{8}
   -\frac{3}{4} l_Z \Bigg] + a_b^2 \Bigg[ -\frac{1}{8} +\frac{3}{4} l_Z \Bigg]
   \nonumber \\
&&{} + \frac{M_Z^2}{M_H^2} a_b^2 \Bigg[ -1 -3 l_Z \Bigg] +\frac{M_W^2}{M_H^2}
\frac{1}{s_w^2} \Bigg[ -\frac{1}{8} - \frac{3}{8} l_W \Bigg]
%
%
  +\frac{M_W^2}{M_t^2} \frac{1}{s_w^2} \Bigg[ \frac{3}{32} +\frac{3}{32}
l_{Wt} \Bigg] \nonumber \\
%
%
&&{} +\frac{M_W^4}{M_t^4} \frac{1}{s_w^2} \Bigg[ \frac{3}{32} +\frac{3}{16}
l_{Wt} \Bigg]
%
%
 +\frac{M_W^6}{M_t^6} \frac{1}{s_w^2} \Bigg[ \frac{3}{32} +\frac{9}{32}
l_{Wt} \Bigg] \nonumber \\
%
%
&&{} +\frac{M_W^8}{M_t^8} \frac{1}{s_w^2} \Bigg[ \frac{3}{32} +\frac{3}{8}
l_{Wt} \Bigg]
%
%
 +\frac{M_W^{10}}{M_t^{10}} \frac{1}{s_w^2} \Bigg[ \frac{3}{32}
+\frac{15}{32} l_{Wt} \Bigg] \nonumber \\
%
%
&&{} +\frac{M_b^2}{M_W^2} \Bigg\{ \frac{1}{s_w^2} \Bigg[
\frac{11}{192} -\frac{1}{16} l_b +\frac{1}{32} l_t
+\frac{1}{32} l_Z +\frac{3}{32} l_H \Bigg]
+\frac{M_W^2}{M_Z^2} v_b^2 \Bigg[ -\frac{2}{3} -l_{bZ} \Bigg] \Bigg\} \nonumber\\ 
&&{}
+\frac{\alpha\alpha_s}{\pi^2} C_F
%
%
 \Bigg\{ \frac{M_t^4}{M_W^2 M_H^2} \frac{N_c}{s_w^2} \Bigg[ -\frac{1}{8}
 - \frac{3}{16} l_b - l_t -\frac{3}{16} l_b l_t - \frac{3}{8} l_t^2 \Bigg]
\nonumber\\
&&{} +\frac{M_t^2}{M_W^2} \frac{1}{s_w^2} \Bigg[ -\frac{13}{64} + \frac{3}{16}
  \zeta (2) +\frac{15}{256} l_b +\frac{3}{32} l_t +\frac{9}{128} l_b l_t
  +\frac{9}{128} l_t^2 \Bigg] \nonumber\\
&&{} +\frac{M_H^2}{M_W^2} \frac{1}{s_w^2}(1+l_H)
\Bigg[\frac{3}{32}+\frac{9}{128}l_b\Bigg]
%
%
+\frac{1}{s_w^2} \Bigg[ -\frac{111}{256} +\frac{3}{16} \zeta (2)
+\frac{9}{64} l_t \Bigg] \nonumber \\
&&{} + Q_b^2 \Bigg[ \frac{7}{64} +6\zeta (2) \ln 2 -\frac{15}{4} \zeta (2)
-\frac{3}{2} \zeta (3)
+\frac{21}{16} l_b +\frac{9}{16} l_b^2 \Bigg] \nonumber\\
&&{} +v_b^2 \Bigg[ \frac{23}{64} +\frac{15}{32} l_b
+\frac{9}{16} l_Z +\frac{9}{16} l_b l_Z \Bigg] + a_b^2 \Bigg[ \frac{55}{64}
+\frac{3}{32} l_b +\frac{9}{16} l_Z -\frac{9}{16} l_b l_Z \Bigg] \nonumber\\
&&{} +\frac{M_Z^2}{M_H^2} \frac{1}{s_w^2 c_w^2}
(1+3l_Z)\Bigg[ \frac{1}{16}+\frac{3}{64} l_b\Bigg]
+\frac{M_W^2}{M_H^2} \frac{1}{s_w^2} 
(1+3l_W)\Bigg[ \frac{1}{8} +\frac{3}{32} l_b\Bigg] \nonumber \\
%
%
&&{} +\frac{M_W^2}{M_t^2} \frac{1}{s_w^2} \Bigg[ -\frac{81}{128} +\frac{21}{64}
\zeta (2) -\frac{9}{128} l_b -\frac{3}{128} l_{Wt} -\frac{9}{128} l_b l_{Wt}
\Bigg] \nonumber\\
%
%
&&{} +\frac{M_W^4}{M_t^4} \frac{1}{s_w^2} \Bigg[ -\frac{305}{384} +\frac{15}{32}
\zeta (2) -\frac{9}{128} l_b -\frac{5}{64} l_{Wt} -\frac{9}{64} l_b l_{Wt}
\Bigg] \nonumber\\
%
%
&&{} +\frac{M_W^6}{M_t^6} \frac{1}{s_w^2} \Bigg[ -\frac{377}{384} +\frac{39}{64}
\zeta (2) -\frac{9}{128} l_b -\frac{5}{64} l_{Wt} -\frac{27}{128} l_b l_{Wt}
\Bigg] \nonumber\\
%
%
&&{} +\frac{M_W^8}{M_t^8} \frac{1}{s_w^2} \Bigg[ -\frac{22669}{19200}
+\frac{3}{4} \zeta (2) -\frac{9}{128} l_b
-\frac{13}{320} l_{Wt} -\frac{9}{32} l_b l_{Wt} \Bigg] \nonumber\\
%
%
&&{} +\frac{M_W^{10}}{M_t^{10}} \frac{1}{s_w^2} \Bigg[ -\frac{106451}{76800}
+\frac{57}{64} \zeta (2) -\frac{9}{128} l_b
+\frac{33}{1280} l_{Wt} -\frac{45}{128} l_b l_{Wt} \Bigg] \nonumber\\
%
%
&&{} +\frac{M_b^2}{M_W^2} \Bigg\{
\frac{1}{s_w^2} \Bigg[ \frac{77}{216} +\frac{\zeta (2)}{24}
-\frac{229}{2304} l_b -\frac{89}{1152} l_t +\frac{19}{1152} l_Z +\frac{3}{128}
l_H \nonumber\\
&&{}-\frac{9}{128} l_b l_t +\frac{17}{384} l_b l_Z -\frac{27}{128} l_b l_H
+\frac{7}{192} l_b^2 +\frac{3}{128} l_t^2 -\frac{13}{384} l_Z^2
 +\frac{9}{128} l_H^2 \Bigg]\nonumber\\
&&{}+\frac{M_W^2}{M_Z^2} v_b^2 \Bigg[ -\frac{25}{18} +\zeta (2) -\frac{8}{9}
l_b  +\frac{25}{18} l_Z +\frac{13}{12} l_b l_Z -\frac{11}{12} l_b^2 -\frac{1}{6}
l_Z^2 \Bigg] \nonumber\\
&&{}
 +\frac{M_W^2}{M_t^2} \frac{1}{s_w^2} \Bigg[ -\frac{13}{96} +\frac{\zeta
  (2)}{16} +\frac{5}{96} l_{bt} \Bigg] 
\Bigg\} 
\Bigg\}+\cdots.
\end{eqnarray}
At $\mathcal{O}(\alpha\alpha_s)$, the term quadratic in $M_t$ agrees with the
result of Ref.~\cite{Kwiatkowski:1995vy}.

We conclude this section by presenting the relationship through 
$\mathcal{O}(\alpha\alpha_s)$ between the $\mathrm{\overline{MS}}$ and pole
masses of the first four quark flavours.
It reads:
\begin{eqnarray}
 \frac{\overline{m}_q(\mu)}{M_q} &=& 1 +\frac{\alpha_s}{\pi} C_F \Bigg\{
-1
 -\frac{3}{4} l_q \Bigg\} +\frac{\alpha}{\pi} \Bigg\{ \frac{M_t^4}{M_W^2
   M_H^2}\frac{N_c}{s_w^2} \Bigg[ \frac{1}{4} +\frac{1}{4} l_t \Bigg]
 +\frac{M_H^2}{M_W^2} \frac{1}{s_w^2} \Bigg[ -\frac{3}{32} -\frac{3}{32}
l_H
 \Bigg]
\nonumber\\
&&{}  - \frac{1}{s_w^2}\frac{3}{32} + Q_q^2 \Bigg[ -1 -\frac{3}{4} l_q
\Bigg] +
v_q^2 \Bigg[ -\frac{5}{8}
   -\frac{3}{4} l_Z \Bigg] + a_q^2 \Bigg[ -\frac{1}{8} +\frac{3}{4} l_Z
\Bigg]
\nonumber\\
&&{} + \frac{M_Z^2}{M_H^2} a_q^2 \Bigg[ -1 -3 l_Z \Bigg]
+\frac{M_W^2}{M_H^2}
\frac{1}{s_w^2} \Bigg[ -\frac{1}{8} - \frac{3}{8} l_W \Bigg] \Bigg\}
\nonumber\\
&&{}+\frac{\alpha\alpha_s}{\pi^2} C_F
 \Bigg\{ \frac{M_t^4}{M_W^2 M_H^2} \frac{N_c}{s_w^2} \Bigg[ -\frac{1}{8}
 - \frac{3}{16} l_q - l_t -\frac{3}{16} l_q l_t - \frac{3}{8} l_t^2 \Bigg]
\nonumber\\
&&{}+\frac{M_H^2}{M_W^2} \frac{1}{s_w^2}(1+l_H)
\Bigg[\frac{3}{32}+\frac{9}{128}l_q\Bigg]
+\frac{1}{s_w^2} \Bigg[ \frac{39}{256} +\frac{9}{128} l_q
+\frac{9}{64} l_W \Bigg]
\nonumber\\
&&{}+ Q_q^2 \Bigg[ \frac{7}{64} +6\zeta (2) \ln 2 -\frac{15}{4} \zeta (2)
-\frac{3}{2} \zeta (3)+\frac{21}{16} l_q +\frac{9}{16} l_q^2 \Bigg]
\nonumber\\
&&{}+v_q^2 \Bigg[ \frac{23}{64} +\frac{15}{32} l_q
+\frac{9}{16} l_Z +\frac{9}{16} l_q l_Z \Bigg] + a_q^2 \Bigg[
\frac{55}{64}
+\frac{3}{32} l_q +\frac{9}{16} l_Z -\frac{9}{16} l_q l_Z \Bigg]
\nonumber\\
&&{}+\frac{M_Z^2}{M_H^2} a_q^2
(1+3l_Z)\Bigg[ 1+\frac{3}{4} l_q\Bigg]
+\frac{M_W^2}{M_H^2} \frac{1}{s_w^2}
(1+3l_W) \Bigg[ \frac{1}{8} +\frac{3}{32}l_q\Bigg] \Bigg\}
+\cdots.
\end{eqnarray}


\section{Summary and conclusion\label{secsum}}

We calculated the $\mathcal{O}(\alpha\alpha_s)$ corrections to the
relationships between the $\mathrm{\overline{MS}}$ Yukawa couplings and the
pole masses of the first five quark flavours in the SM.
We demonstrated that these relationships are gauge-parameter independent and
free of tadpole contributions.

The method of asymptotic expansion was used in the calculation, and the
results were found as polynomials in small mass ratios with coefficients that
contain logarithms of these mass ratios.
The goodness of this method was tested in two ways.
On the one hand, our $\mathcal{O}(\alpha)$ expression for $\delta_b(M_b)$ was
found to reproduce the analytic result of Ref.~\cite{Hempfling:1994ar}
extremely well.
On the other hand, our $\mathcal{O}(\alpha\alpha_s)$ result turned out to
converge very fast as higher powers of $M_W^2/M_t^2$ were included.
These two observations reassure us of the reliability of our result, which
is equivalent to the exact analytic result for all practical purposes.

As for the phenomenological significance of our results, our numerical
analysis revealed that, at $M_H=100$~GeV, the $\mathcal{O}(\alpha\alpha_s)$
contribution to $\delta_b(M_b)$ amounts to roughly one half of the
$\mathcal{O}(\alpha)$ one.
These two contributions depend very differently on the value of $M_H$, and
they even cross over at $M_H\approx500$~GeV.
Obviously, the $\mathcal{O}(\alpha\alpha_s)$ contribution cannot be neglected
against the $\mathcal{O}(\alpha)$ one.

We also calculated the $\mathcal{O}(\alpha\alpha_s)$ corrections to the
relationships between the $\mathrm{\overline{MS}}$ and pole masses of the
first five quark flavours.
We showed that these relationships are gauge-parameter independent if the
tadpole contributions are properly included in the contributing self-energies. 


\end{document}